\documentclass[conference]{IEEEtran}
\usepackage{subfig}
\usepackage{array}
\usepackage{graphicx}
\usepackage{amsmath}
\usepackage{algorithmic}
\usepackage{algorithm}
\usepackage[bookmarks=false]{hyperref}
\usepackage{doi}
\usepackage{url}

\usepackage{setspace}
\usepackage{enumitem}
\usepackage{graphicx,lipsum}
\usepackage{caption}
\usepackage{tikz}

\newcolumntype{P}[1]{>{\centering\arraybackslash}p{#1}} 
\newcolumntype{M}[1]{>{\centering\arraybackslash}m{#1}} 
\usepackage[compact]{titlesec}

\newcommand\copyrighttext{%
  \footnotesize \copyright~2015 IEEE. Personal use of this material is permitted. Permission from IEEE must be obtained for all other uses, in any current or future media, including reprinting/republishing this material for advertising or promotional purposes, creating new collective works, for resale or redistribution to servers or lists, or reuse of any copyrighted component of this work in other works.}
\newcommand\copyrightnotice{%
\begin{tikzpicture}[remember picture,overlay]
\node[anchor=south,yshift=10pt] at (current page.south) {\fbox{\parbox{\dimexpr\textwidth-\fboxsep-\fboxrule\relax}{\copyrighttext}}};
\end{tikzpicture}%
}
\begin{document}
\bibliographystyle{IEEEtran}
%

\title{Reliable Prediction of Channel Assignment Performance in Wireless Mesh Networks}

\author{\IEEEauthorblockN{Srikant Manas Kala, Ranadheer Musham, M Pavan Kumar Reddy, and Bheemarjuna Reddy Tamma}
\IEEEauthorblockA{ Indian Institute of Technology Hyderabad, India\\
Email: [cs12m1012, cs12b1026, cs12b1025, tbr]@iith.ac.in}}

\maketitle
\copyrightnotice
\begin{abstract}
The advancements in wireless mesh networks (WMN), and the surge in multi-radio multi-channel (MRMC) WMN deployments have spawned a multitude of network performance issues. These issues are intricately linked to the adverse impact of endemic interference. Thus, interference mitigation is a primary design objective in WMNs. Interference alleviation is often effected through efficient channel allocation (CA) schemes which fully utilize the potential of MRMC environment and also restrain the detrimental impact of interference. However, numerous CA schemes have been proposed in research literature and there is a lack of CA performance prediction techniques which could assist in choosing a suitable CA for a given WMN. In this work, we propose a reliable interference estimation and CA performance prediction approach. We demonstrate its efficacy by substantiating the CA performance predictions for a given WMN with experimental data obtained through rigorous simulations on an ns-3 802.11g environment.
\end{abstract}

\section{Introduction}
Multi-radio multi-channel wireless mesh networks (MRMC WMNs) are expected to significantly reduce the dependence on wired network infrastructure owing to the availability of low-cost commodity IEEE 802.11 hardware, ease of scalability, and flexibility in deployment. MRMC WMNs offer reliable connectivity by leveraging the inherent redundancy in the underlying mesh topology framework. 
This is facilitated by multiple-hop transmissions which relay the data traffic seamlessly between source-destination pairs where a direct communication can not be established \cite{6Akyildiz}. 
However, the broadcast nature of wireless transmissions is synonymous with \textit{link conflicts} spawned by WMN radios which are located within each other's interference range and are concurrently active on an identical channel. The lucrative features of MRMC WMNs \emph{viz.}, enhanced capacity, seamless connectivity, and reduced latency are diminished by the adverse impact of interference. \textit{Conflict graphs} (CGs) are invariably used to represent these interference complexities in a WMN. A CG models the wireless links in a WMN as vertices and edges between these vertices represent potential link conflicts \cite{22Ramachandran}. Interference alleviation in WMNs is primarily accomplished through an efficient channel assignment (CA) to the radios. Thus, the intensity of interference affecting a WMN is the characteristic of the implemented CA scheme, as it is responsible for reigning in the endemic interference. However, the CA problem is an NP-Hard problem \cite{NPcomplete} and several CA schemes have 
been proposed in literature \cite{Ding}
 which employ numerous concepts and heuristic approaches to mitigate the impact of interference in a WMN.

\section{Motivation and Related Research Work}

Estimation of interference, its alignment and cancellation are well known NP-Hard problems \cite{Alignment}. Numerous research endeavors have tried to address the interference alignment and cancellation at the physical layer \emph{e.g.}, in \cite{Cancel}, authors employ the \textit{soft interference cancellation} technique. Impact of interference on multi-hop wireless networks has also been rigorously studied, maximum achievable \textit{network capacity} being the primary focus of these studies. In the landmark work \cite{GuptaKumar}, authors demonstrated that in a wireless network consisting of $n$ randomly placed identical nodes, where each node is communicating with another, the maximum achievable throughput per node is $\Theta(1/\sqrt{n \log{} n})$. In \cite{Capacity}, authors estimate the network capacity of an arbitrary wireless network by employing a realistic \textit{signal to interference plus noise ratio} (SINR) model to account for the interference. 
Authors in \cite{Impact} assess the impact of interference in multi-hop mesh networks by proposing an upper bound on the achievable network capacity, under the constraints of specific physical location of wireless nodes and a particular traffic load.
The concept of \textit{interference degree} (ID) is often used in solutions to the resource allocation \cite{TID1}, scheduling \cite{TID2}, and CA problems \cite{Arunabha}, with the intertwined objectives of minimizing the prevalent interference and optimizing the WMN performance. ID of a wireless link in a WMN denotes the number of links in its close proximity which can potentially interfere with it \emph{i.e.}, disrupt a transmission on the given link. Total interference degree (TID) of a WMN is obtained by halving the sum of ID of all links in the WMN. In our previous work \cite{Manas}, we highlight that TID is only an approximate measure of the intensity of interference but not a dependable CA performance prediction metric. Further in \cite{Manas3}, we propose a fresh characterization of interference, attributing three dimensions, namely, \textit{statistical, spatial and temporal}, to the interference prevalent in wireless networks. Based on this characterization, a statistical \textit{Channel 
Distribution Across Links} (CDAL) algorithm is suggested which identifies the link-count for each channel \emph{i.e.}, the 
number of links in the wireless network that have been allocated that particular channel. It then computes a statistical metric CDAL$_{cost}$, which is a measure of equitable distribution of channels across wireless links. Further, CDAL$_{cost}$ is demonstrated to be a more reliable estimation metric than TID, at a lesser computational cost. 

Thus, apart from TID estimate and CDAL$_{cost}$ there is an absence of alternate metrics in research literature, which can be employed as well founded theoretical benchmarks for comparison and prediction of CA performance. In this study, we further bridge that gap by using the interference characterization model of \cite{Manas3} to engineer a \textit{spatio-statistical} interference estimation and CA predication scheme.
\section{Problem Definition}
Let $G=(V,E)$ represent an arbitrary MRMC WMN comprising of $n$ nodes, where $V$ denotes the set of all nodes and $E$ denotes the set of wireless links in the WMN. Each node $i$ is equipped with a random number of identical radios $R_i$, and is assigned a list of channels $Ch_i$ from the set of available channels $Ch$. A reliable theoretical interference estimate needs to be devised to predict with high confidence, the efficient CA schemes that ought to be selected for $G$ from the available set of CA schemes. 

\section{Interference Estimation \& CA Performance Prediction}
The proposed algorithm adopts a comprehensive two dimensional \textit{spatio-statistical} view of prevalent interference. The \textit{spatial} dimension concentrates on the link conflicts which are spawned due to spatial proximity of radios, while the \textit{statistical} dimension is concerned with a proportional distribution of channels across wireless links in a WMN. 
\subsection{Inadequacy of Statistical Interference Estimation}
Leveraging the statistical aspects of endemic interference offers a sound estimation metric in CDAL$_{cost}$. However, there is a lacuna in the CDAL algorithm that it accounts for only a single dimension of interference. We now highlight this limitation of CDAL estimation.

 \begin{figure}[htb!]
                \centering
                \includegraphics[width=8.5cm, height=3cm]{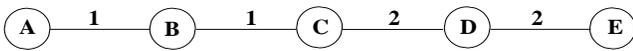}
                \caption{Limitation of CDAL estimation}
                \label{stat}
        \end{figure}

Figure~\ref{stat} depicts channel allocations in a $5$ node MRMC WMN under two CA schemes, CA$_X$ and CA$_Y$. Two non-overlapping channels \textit{(1, 2)} are available to the CA schemes. The channel allocations to the link quartet \textit{(AB, BC, CD, DE)} under CA schemes CA$_X$ and CA$_Y$ are, (1, 2, 1, 2) and (1, 1, 2, 2), respectively. For a smooth discourse, we assume a \textit{Transmission : Interference} range of 1:1 \emph{i.e.}, only the transmissions over adjacent links interfere. It can be inferred that the two channel allocations are statistically alike \emph{i.e.}, link-counts of both the channels are identical under both CA schemes. But the CA schemes differ in terms of spatial distribution of links in the network. The spatial features of CA$_X$ guarantee a minimal interference scenario, as adjacent links transmit over non-overlapping channels. In sharp contrast, CA$_Y$ leads to a high interference scenario where adjacent links (AB \& BC) and (CD \& DE) operate over identical channels and 
cause link conflicts. 
The CDAL algorithm is oblivious to these spatial characteristics and assigns the two CA schemes the same CDAL$_{cost}$. This causes an erroneous prediction and comparison of CA performance, which renders CDAL estimate less accurate. However, it forms the theoretical foundation for a more efficient estimation technique which we propose next.    

\renewcommand{\algorithmicrequire}{\textbf{Input:}}
\renewcommand{\algorithmicensure}{\textbf{Output:}}
\begin{algorithm}[htb!] 
\caption{Cumulative X-Link-Set Weight}
\label{xlink}
\begin{algorithmic}[1]
{\fontsize{9}{10}
\REQUIRE $G = (V,E)$, $R_i (i \in V)$, $CA = \{(R_i,CS), i \in V\}$,\\ $CS =\{1, 2,...M\}$ \\
\textit{Notations} $:$ $G$ $\leftarrow$  WMN Graph, $R_i$ $\leftarrow$ Radio-Set,\\
		   $CS$ $\leftarrow$ Available Channel Set, $CA$ $\leftarrow$ Channel Assignment
\ENSURE $CXLS_{wt}$ \\
\line(1,0){236}
\FOR {$i \in V$}
\STATE Determine $Ch_i$ and $Adj_i$ 
\COMMENT {$Ch_i$  : Set of channels allocated to the radios at node $i$ in $G$.  $Adj_i$ : Set of nodes adjacent to node $i$ in $G$}
\ENDFOR
\FOR {$i \in V$}
\FOR {$j \in Adj_i$}
\STATE $LnSet \leftarrow InsertLn(i,j)$. \COMMENT {$LnSet$ : Set of all possible wireless links in of $G$.}
\STATE $ComCh_{ij} \leftarrow GetComCh(Ch_i, Ch_j,)$ \COMMENT {$ComCh_{ij}$: Set of common channels assigned to radios of $(i \ \& \ j)$}
\STATE $LnChMap \leftarrow InsertLnCh(LnChMap, i, j, ComCh_{ij})$ \COMMENT{$LnChMap$ : Contains Link-Channel mapping.}
\ENDFOR
\ENDFOR
\STATE Let $Transmission \ Range:Interference \ Range = 1:X$ 
\STATE $SXLS  \leftarrow GetAllLinkSets(LnSet,X)$  \COMMENT {$SXLS$ : Set of all X-link-Sets.}
\STATE  $CXLS_{wt}\ \leftarrow \ ProbCompWeight(SXLS,LnChMap, X)$. \COMMENT {Function $ProbCompWeight()$ implements Algorithm~\ref{probwt}}
\STATE{Output the $CXLS_{weight}$}
}
\end{algorithmic}
\end{algorithm}

\subsection{Spatio-Statistical Interference Estimation}
Any theoretical interference estimation scheme can only account for the spatio-statistical aspects of the three dimensional interference estimation problem. An intelligent spatio-statistical scheme will not just factor in the spatial proximity of links, but will also consider the distribution of available channels among the radios, thereby offering an efficient CA performance estimation metric. The algorithm considers the wireless links in a WMN and assigns a \textit{set of links} a certain \textit{weight}, which reflects its resilience to the adverse impact of interference. We call it the \textit{Cumulative X-Link-Set Weight} or $CXLS_{wt}$ algorithm and present it in Algorithm~\ref{xlink}.

We begin the discourse by explaining the term \textit{X-Link-Set}. In a wireless network, the interference range of a radio \emph{i.e.}, the distance over which the signal strength is potent enough to interfere with another signal but unable to successfully deliver data, far exceeds its transmission range \emph{i.e.}, the distance upto which the signal strength of a transmission guarantees a successful data delivery at a receiving radio. The ratio of \textit{Transmission Range : Interference Range} (T:I) in most wireless networks usually lies between 1:2 to 1:4. In the proposed algorithm a T:I of 1:X is considered, where X is a positive integer. The factor $X$ has a great significance in determining the detrimental effect of interference on a link. For example, in Figure~\ref{stat} the channel allocation of CA$_X$ is optimal for a T:I of 1:1, but for 1:2 both CA$_X$ and CA$_Y$ experience the same number of link conflicts. Thus $X$ determines the \textit{impact radius} (IR) of link conflicts, and it 
ought to be taken into consideration while designing an interference estimation algorithm. The CXLS$_{wt}$ algorithm accounts for the impact radius $X$ by considering a set of $X$ consecutive links named the \textit{X-Link-Set} or $XLS$ as the fundamental entity for interference estimation.

The CXLS$_{wt}$ algorithm begins by determining the set of channels assigned to the radios of each node and the adjacency list of each node. Next, all wireless links in the WMN are determined on the basis of transmission range \emph{i.e.}, adjacency of nodes in the graphical representation of the WMN. Further, for each link, the algorithm finds the set of common channels that are assigned to radios of the adjacent nodes which share that particular link. The links are stored in a data structure called $LnSet$ while the channel set associated to the link is mapped to it in $LnChMap$. Further, a set of \mbox{X-Link-Sets} or $SXLS$, is determined by the function $GetAllLinkSets$. $SXLS$ serves as a sample space of fundamental blocks \emph{i.e.}, $XLS$, and the final step entails processing them to generate an interference estimation metric. To each element of this sample space \emph{i.e.}, to every $XLS$, we assign a \textit{weight} which is a measure of its quality. A higher weight signifies a diminished impact 
of interference in the $XLS$, whereas a low weight implies that the $XLS$ is severely degraded by interference. 
\renewcommand{\algorithmicrequire}{\textbf{Input:}}
\renewcommand{\algorithmicensure}{\textbf{Output:}}
\begin{algorithm}[htb!] 
\caption{Computation Of X-Link-Sets Weight}
\label{probwt}
\begin{algorithmic}[1]
{\fontsize{9}{10}
\REQUIRE$SXLS, \ LnChMap, \ X$\\
 $SXLS$ : Set of all X-link-Sets, $X$ : Interference Range,\\
 $LnChMap$ : Set containing Link-Channel mapping.\\
\ENSURE CXLS$_{wt}$\\
\line(1,0){236}
\STATE $CXLS_{wt} \leftarrow 0$
\FOR {$XLS \in SXLS$}
\STATE $XLS_{wt} \leftarrow 0$
\STATE Let {$(Ln_1,Ln_2, \ldots,Ln_X) \in XLS$}
\STATE Let $Ch_1, Ch_2, \ldots Ch_X$ be the set of channels mapped to the corresponding links $Ln_1,Ln_2, \ldots,Ln_X$.
\STATE Assign channels to all $Ln_i$ from the channel-set $Ch_i$
\STATE $TempXLS_{wt} \leftarrow 0$
\STATE $count \leftarrow 0$
\FOR {all \textit{Equally Probable} combinations of channel assignments in $XLS$}
\IF {all $X$ links are assigned identical channels}
\STATE $TempXLS_{wt}\leftarrow 0$
\ELSIF {$X-1$ links are assigned identical channels, $1$ link is assigned an orthogonal channel}
\STATE $TempXLS_{wt} \leftarrow 1$\\
$\vdots$\\
\ELSIF {all $X$ links are assigned non-overlapping channels}
\STATE $TempXLS_{wt} \leftarrow X$
\ENDIF
\STATE $XLS_{wt} \leftarrow XLS_{wt} + TempXLS_{wt}$
\STATE $count \leftarrow count+1$

\ENDFOR
\STATE $XLS_{wt} \leftarrow XLS_{wt}/count$
\STATE $CXLS_{wt} \leftarrow CXLS_{wt} + XLS_{wt}$
\ENDFOR
\STATE Return $CXLS_{wt}$ 
}
\end{algorithmic}
\end{algorithm}

The technique of weight assignment appeals to the spatial characteristics of interference and is described in Algorithm~\ref{probwt} which is implemented in function $ProbCompWeight$. From the $SXLS$, individual $XLSs$ are selected and processed iteratively. An $XLS$ is further split up into its $X$ consecutive constituent links, $Ln_i$ where $i \in (1\ldots X)$. For each $Ln_i$, the set of channels $Ch_i$ associated to it are retrieved. Next, for each $XLS$ all possible combinations of channel assignments to $Ln_i$ from their respective $Ch_i$ are generated. The motivating principle for considering all possible channel allocation variations for an $XLS$ is the same as probabilistic selection of channels in the CDAL algorithm. The channel selection for a link being a temporal characteristic, we account for this dynamism and randomness in the system by considering all the variations as equally probable. Thus, for every channel allocation pattern, the algorithm assigns a weight based on the spatial proximity 
of links. The final weight for an $XLS$ is the average of all of its variations. Within an $XLS$ channel allocation instance, if all of the X links are assigned an identical channel, the weight assigned to the $XLS$ is 0 which is the minimum weight. This scenario defines a maximal interference scenario \emph{i.e.}, every link within the XLS interferes with every other link as the impact radius of $X$ spans the entire $XLS$. Further, if (X-1) links are assigned conflicting channels and 1 link operates on a non-conflicting channel, the weight assigned to the $XLS$ is $1$. For, (X-2) links operating on overlapping channels and 2 links on non-overlapping channels, the weight assigned is 2. Finally, if all the $X$ links are assigned orthogonal channels, which is the minimal interference scenario, a maximum weight of X is assigned to the $XL$S instance. After all of the $XLS$ weights ($XLS_{wt}$) are computed, the algorithm sums them together to generate the final metric for the CA which is the CXLS$_{wt}$. 
 It is noteworthy that a link may be a part of multiple $XLSs$, and will contribute to the weight assignment in each one of them. Hence, the algorithm takes into account all interference scenarios that may arise within a WMN. Further, generating a sample space consisting of $XLS$, assigning each sample a weight, and deriving the metric by a sum of the weights of entire sample space are the statistical features of the CXLS$_{wt}$ algorithm. 
\subsection{Time Complexity of CXLS$_{wt}$ Algorithm}
For an arbitrary MRMC WMN graph $G=(V,E)$, comprising of $n$ nodes and $m$ identical radios installed on every node, the upper-bound on algorithmic complexity of the CXLS$_{wt}$ algorithm can be determined to be O($n\textsuperscript3 m\textsuperscript2$). The $SXLS$ computation incurs an algorithmic cost of O($n\textsuperscript3$) and the cardinality of the set \emph{i.e}, the number of $XLSs$ in the set has an upper bound of O($n\textsuperscript2$). Further, for each $XLS$ the weight is computed by the function $ProbCompWeight()$ by processing each individual link in the $XLS$. This step is the most computationally intensive in the algorithm and has a worst-case complexity of O($n\textsuperscript3 m\textsuperscript2$).

In comparison, TID and CDAL$_{cost}$ estimations have a worst-case algorithmic complexity of O($n\textsuperscript2 m\textsuperscript3$) and O($n\textsuperscript2 m\textsuperscript2$), respectively \cite{Manas3}. Since in any WMN deployment, number of nodes far exceeds the number of radios installed on a node \emph{i.e.}, \textit{n $>>$ m}, CXLS$_{wt}$ estimation requires more computational resources than the other two estimation schemes. However, the results will demonstrate that this slight increase in complexity is a small cost to pay for significantly improved accuracy levels.
\section{Simulations, Results and Analysis}
We now subject the proposed interference estimation algorithm to prove its efficacy in prediction of CA performance in WMNs.
\subsection{Simulation Parameters}
We perform exhaustive simulations in ns-3 \cite{NS-3} to record the performance of CAs in a $5\times5$ grid WMN.  A WMN of grid layout is ideal for evaluating CA efficiency as it outperforms random WMN deployments in terms of metrics such as access-tier coverage area, back-haul connectivity, fairness in channel allocation, and mesh capacity \cite{Grid}.
The simulation parameters are presented in Table~\ref{sim}. Each multi-hop traffic flow transmits a datafile from the source to the destination. TCP and UDP are the underlying transport layer protocols which are implemented through the inbuilt ns-3 models of \textit{BulkSendApplication} and \textit{UdpClientServer}. TCP simulations offer the \textit{aggregate network throughput} while UDP simulations determine the \textit{packet loss ratio} and the \textit{mean delay}. We equip each node in the grid WMN with $2$ identical radios and CA schemes have $3$ orthogonal channels at their disposal. 
\subsection{Test Scenarios}
Multi-hop data flows are an intrinsic feature of WMNs. To gauge the detrimental impact of the endemic interference we design four high traffic test-cases by activating multiple concurrent multi-hop flows. Test scenarios in the grid WMN include a combination of \textit{4-hop flows} from the first node of a row or column to the last node of that particular row or column, and \textit{8-hop flows} which are established between the diagonally opposite nodes placed at the corners of the grid. From various combinations of these two categories of multi-hop flows, four test scenarios are designed which are subjected to both TCP and UDP simulations. They comprise of the following number of concurrent flows which are activated simultaneously in the $25$ node grid : \\ 
\quad(i) \quad 5 \quad   (ii) \quad 8 \quad    (iii) \quad 10 \quad    (iv) \quad 12.
\begin{table} [h!]
\caption{ns-3 Simulation Parameters}
   \center 
\begin{tabular}{|p{5cm}|p{3cm}|}
\hline
\bfseries
 Parameter&\bfseries Value \\ [0.2ex]
 \hline
\hline
Grid Size&$5\times5$   \\
\hline
No. of IEEE 802.11g Radios/Node&2   \\
\hline
Range Of Radios&250 mts   \\
\hline
Available Orthogonal Channels&3 in 2.4 GHz  \\
\hline
Maximum 802.11g PHY Datarate &54 Mbps  \\
\hline
Datafile size &10 MB  \\
\hline
Maximum Segment Size (TCP)&1 KB   \\
\hline
Packet Size (UDP)&1 KB\\
\hline
MAC Fragmentation Threshold&2200 Bytes  \\
\hline
RTS/CTS &Enabled  \\
\hline
Routing Protocol &OLSR    \\
\hline
Loss Model&Range Propagation   \\
\hline
Rate Control&Constant Rate \\
\hline
\end{tabular}
\label{sim}
\end{table}   

\subsection{Selection of CA Schemes} We implement a heterogeneous mix of well-known CA schemes \emph{viz.}, a centralized breadth first traversal approach (BFSCA) \cite{22Ramachandran}, a static maximum clique based algorithm (CLICA) \cite{17Xutao}, a  maximum independent set based scheme (MISCA)  \cite{24Aizaz} and a centralized static CA scheme (CCA) \cite{23Cheng}. We also implement two \textit{radio co-location aware} CA algorithms \emph{viz.}, an optimized independent set based CA scheme (OISCA) and a spatio-statistically designed, elevated interference zone mitigation approach (EIZMCA) \cite{Manas2}. Each of these CA schemes is implemented using two broad based multi-radio multi-channel conflict graph models (MMCGs) \emph{viz.}, the conventional MMCG (C-MMCG) and the enhanced MMCG (E-MMCG) \cite{Manas}. C-MMCG is the traditional way of representing link conflicts, and does not account for radio co-location interference (RCI) prevalent in a wireless network. E-MMCG is a marked improvement over its 
conventional counterpart and adequately represents RCI interference scenarios in its link conflict representation of the WMN. The use of E-MMCG leads to reduced interference levels and improved WMN performance \cite{Manas}, which is also reflected in the results we present in this study. \\
Thus, for all of the above mentioned 6 CA schemes we have two versions, one for each MMCG model, resulting in 12 CAs. In addition, we also implement a grid specific CA scheme (GSCA) for the grid WMN through a crude brute-force approach which permutes through all possible channel allocations in the grid to determine a CA with the minimal TID estimate. It serves as a reference for performance evaluation of the CAs. Finally, we have a total of 13 implementable channel allocations from the 7 CA algorithms. In \cite{Manas3}, the evaluations of CDAL$_{cost}$ estimates were done on a  CA sample set of 9 CAs and in this work the sample set is enlarged to ensure a more comprehensive evaluation. Further, the objective is a quantitative assessment of the CA performances, and not a qualitative analysis of their algorithmic design. Thus, it will suffice to present the results of the simulation exercise and use them as a benchmark to determine the efficiency of the prediction algorithms. To facilitate a smooth referencing,
 we will denote a C-MMCG CA as $CA_C$ and its E-MMCG variant as $CA_E$.
 \begin{figure}[htb!]
                \centering
                \includegraphics[width=7cm, height=5cm]{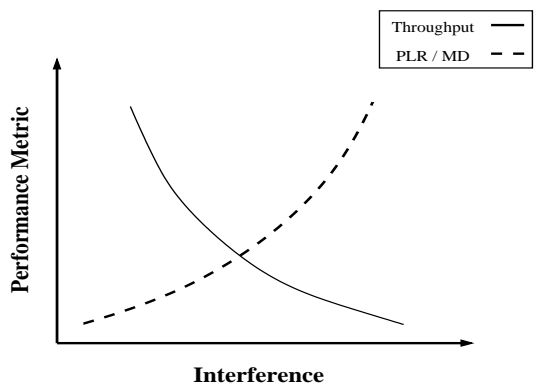}
                \caption{Expected correlation of performance metrics with interference}
                \label{cor}
        \end{figure}
   \begin{figure*}
  \centering%
  \begin{tabular}{cc}
   \subfloat[TID vs Throughput]{\includegraphics[width=.33\linewidth]{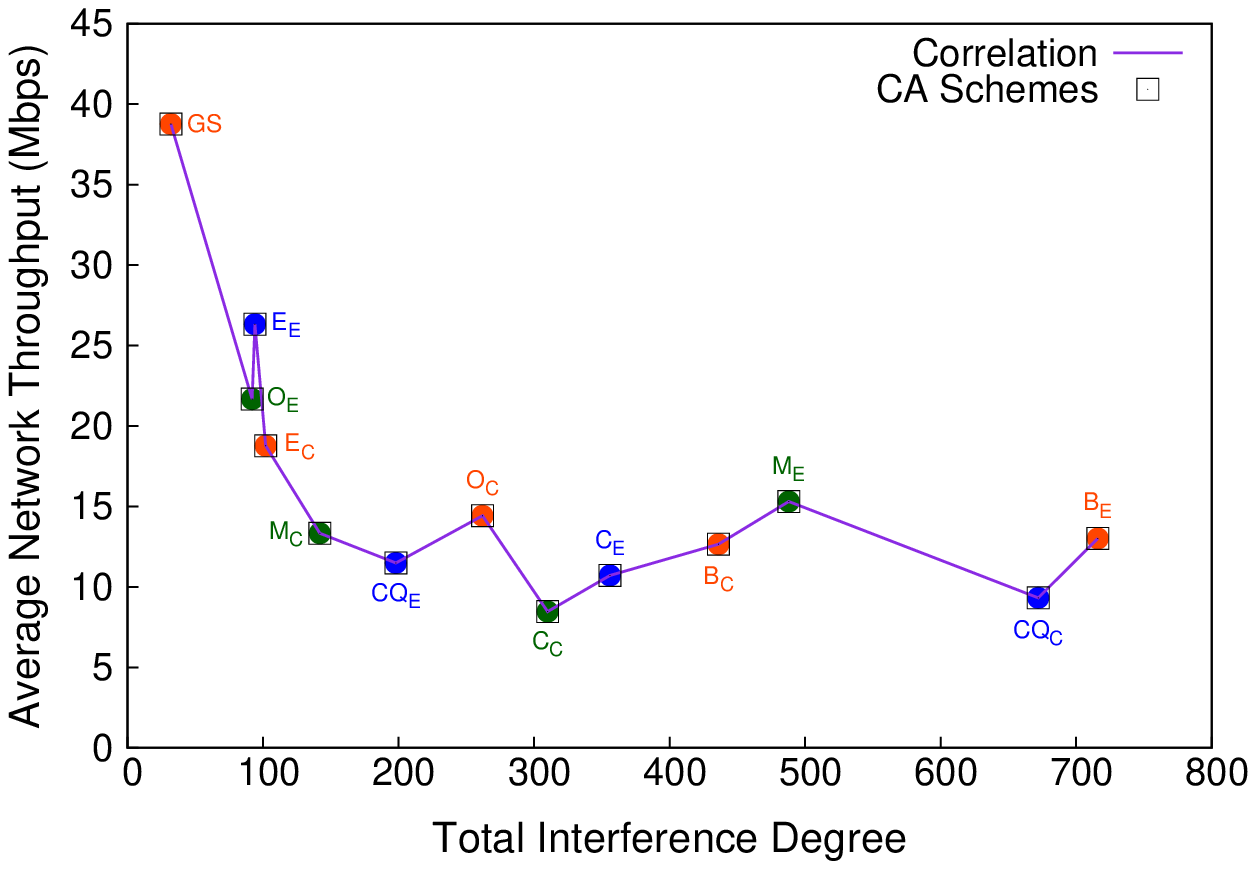}}\hfill%
    \subfloat[CDAL$_{cost}$ vs Throughput]{\includegraphics[width=.33\linewidth]{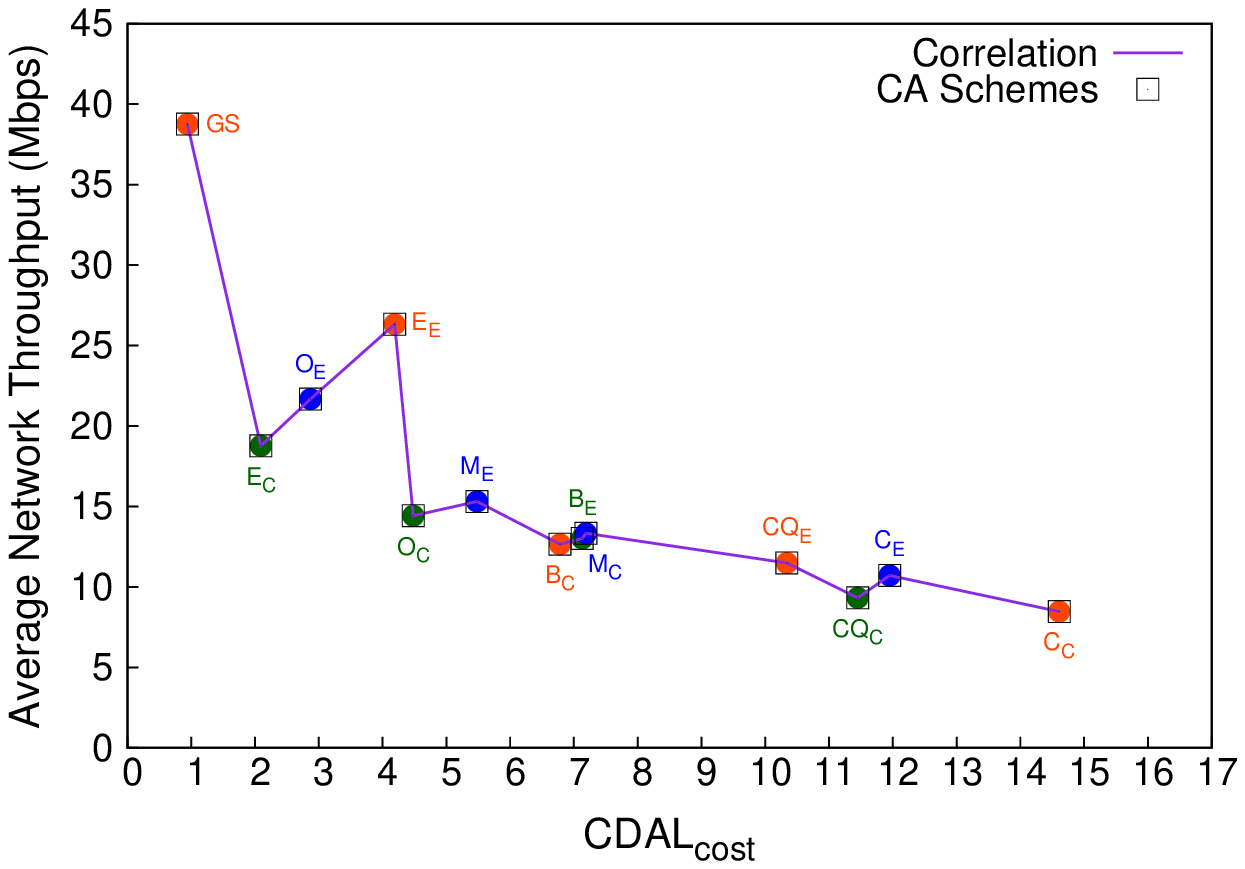}}\hfill%
   \subfloat[CXLS$_{wt}$ vs Throughput] {\includegraphics[width=.33\linewidth]{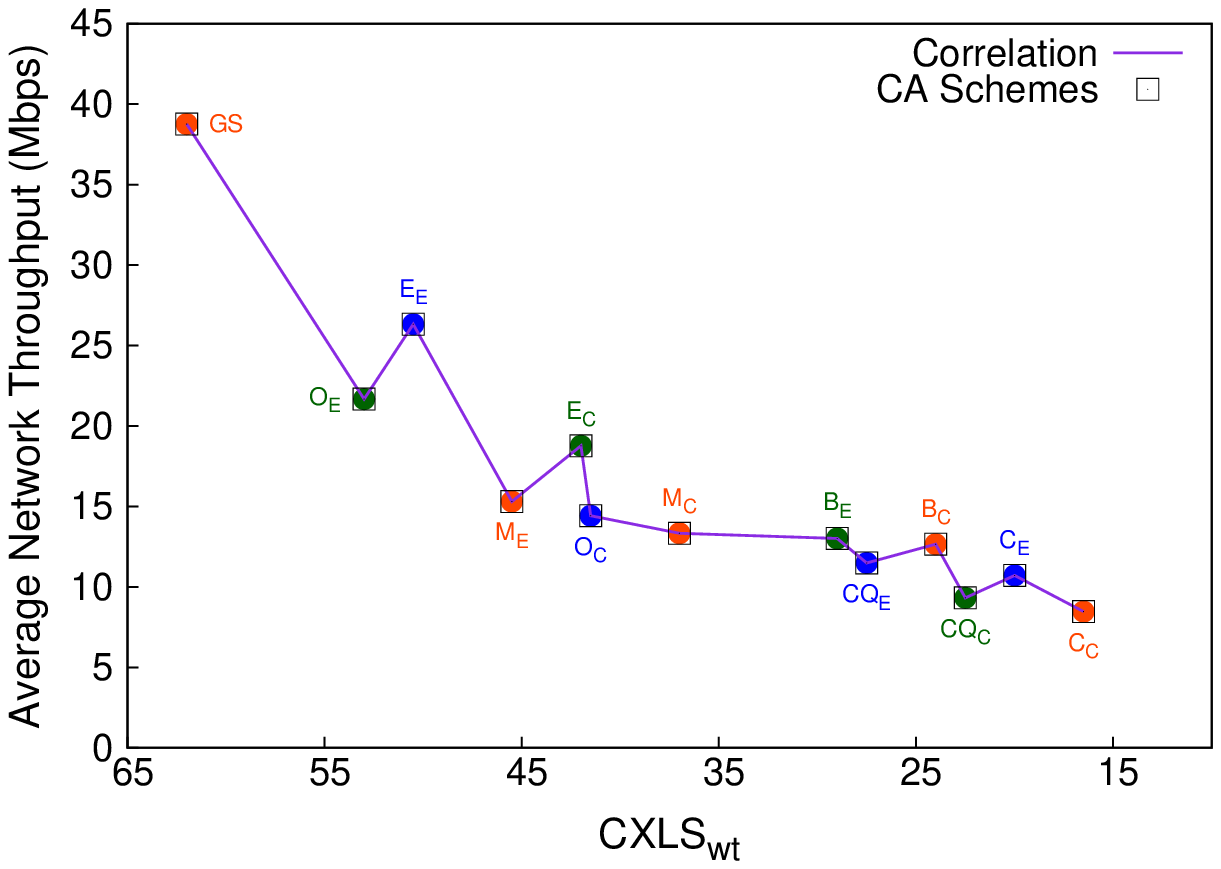}}%
    \end{tabular}
    \caption{Observed correlation of theoretical estimates \& Throughput} 
     \label{cTh}
\end{figure*}

\begin{figure*}
  \centering%
  \begin{tabular}{cc}
   \subfloat[TID vs PLR]{\includegraphics[width=.33\linewidth]{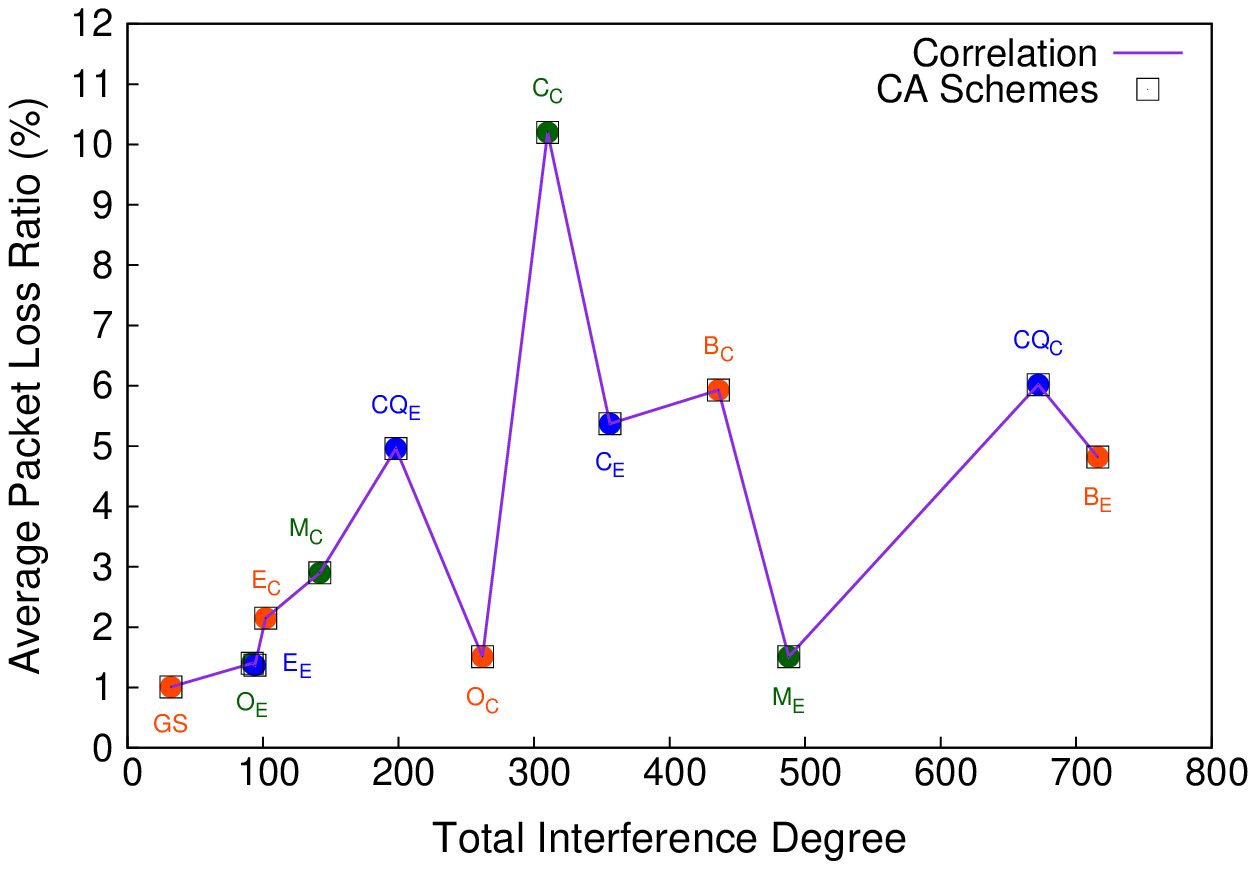}}\hfill%
    \subfloat[CDAL$_{cost}$ vs PLR]{\includegraphics[width=.33\linewidth]{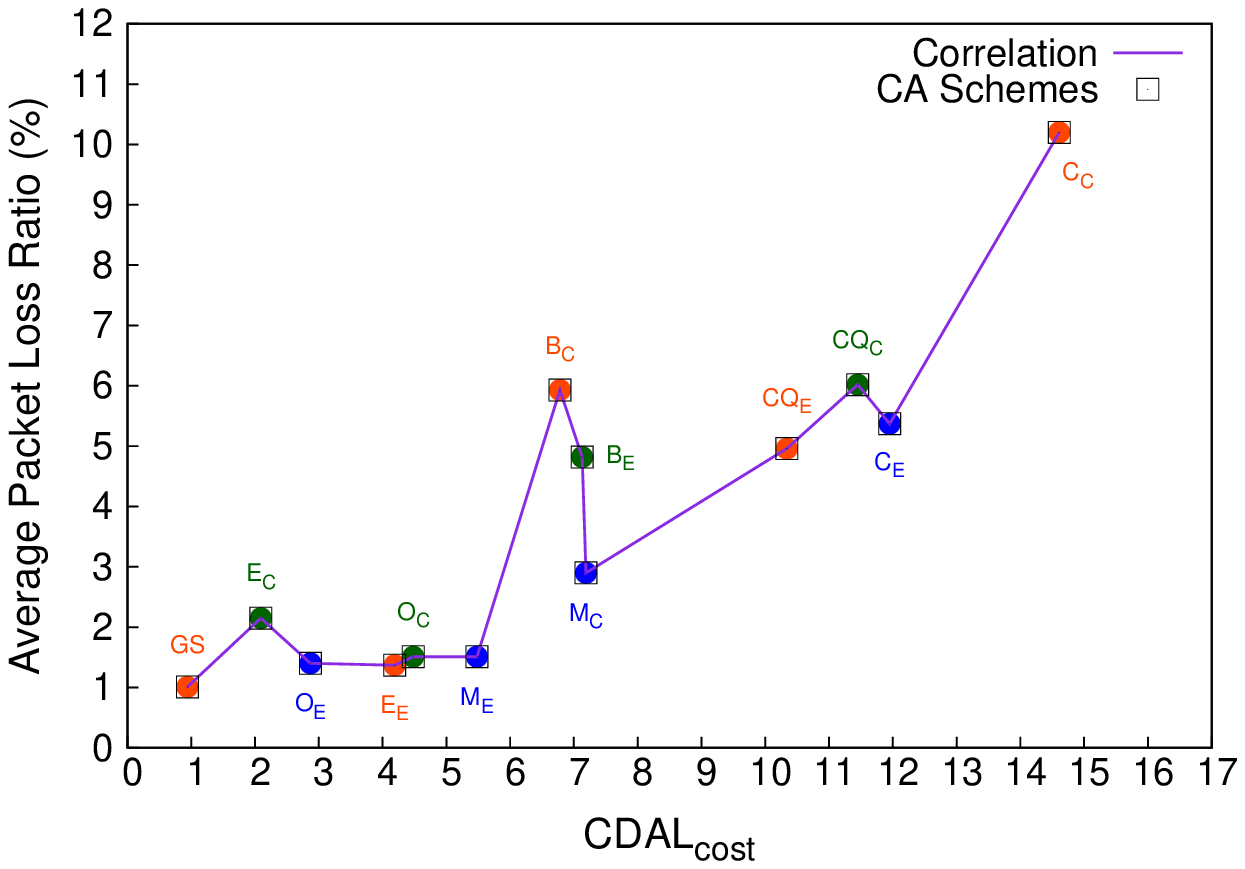}}\hfill%
   \subfloat[CXLS$_{wt}$ vs PLR] {\includegraphics[width=.33\linewidth]{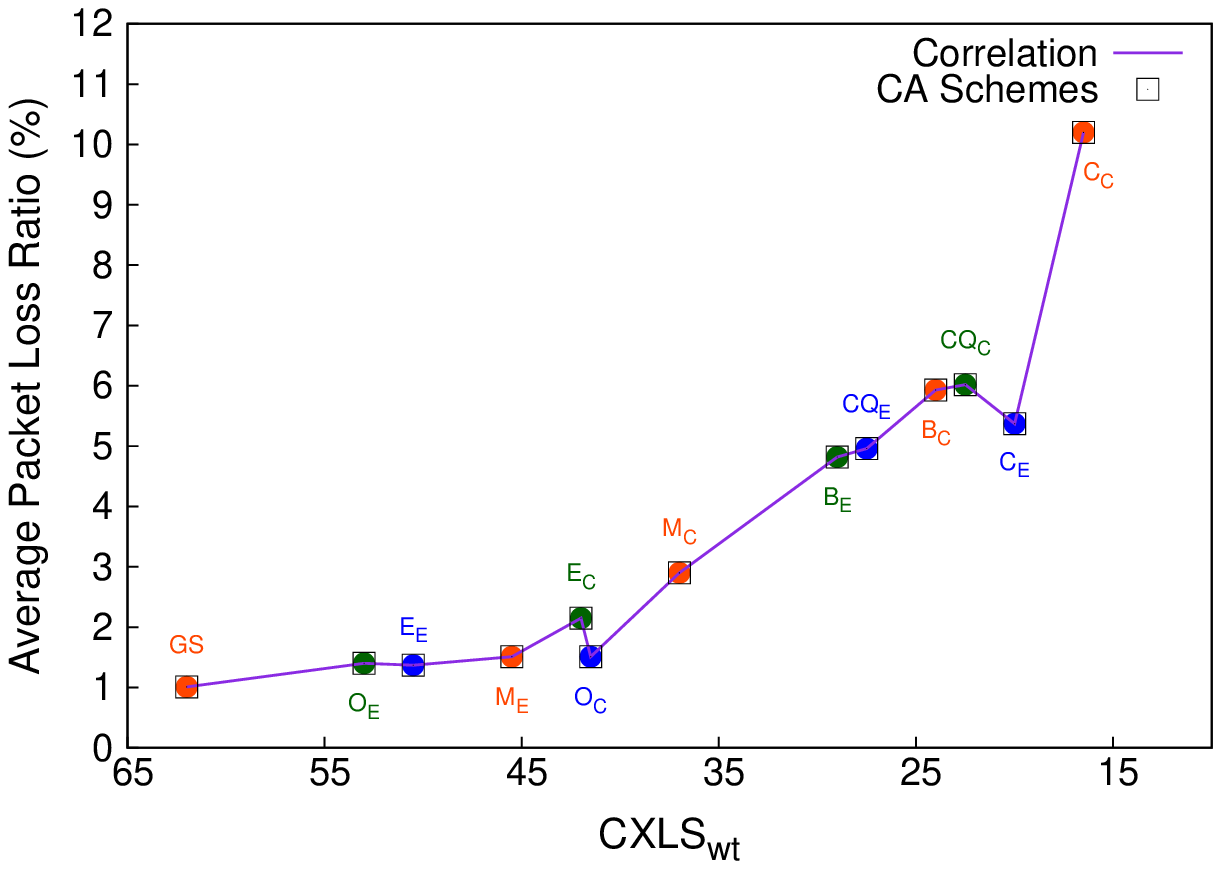}}%
    \end{tabular}
    \caption{Observed correlation of theoretical estimates \& Packet Loss Ratio} 
     \label{cPLR}
\end{figure*}

\begin{figure*}
  \centering%
  \begin{tabular}{cc}
   \subfloat[TID vs MD]{\includegraphics[width=.33\linewidth]{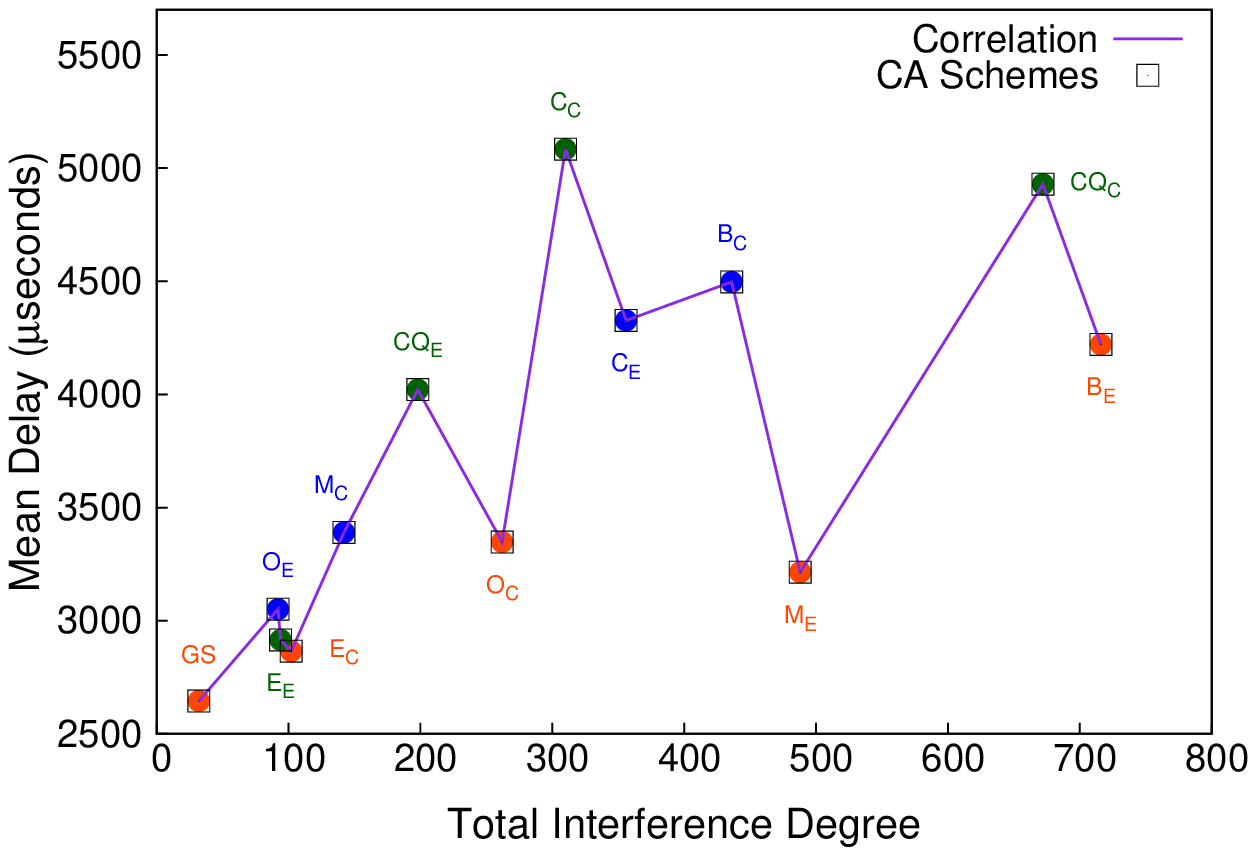}}\hfill%
    \subfloat[CDAL$_{cost}$ vs MD]{\includegraphics[width=.33\linewidth]{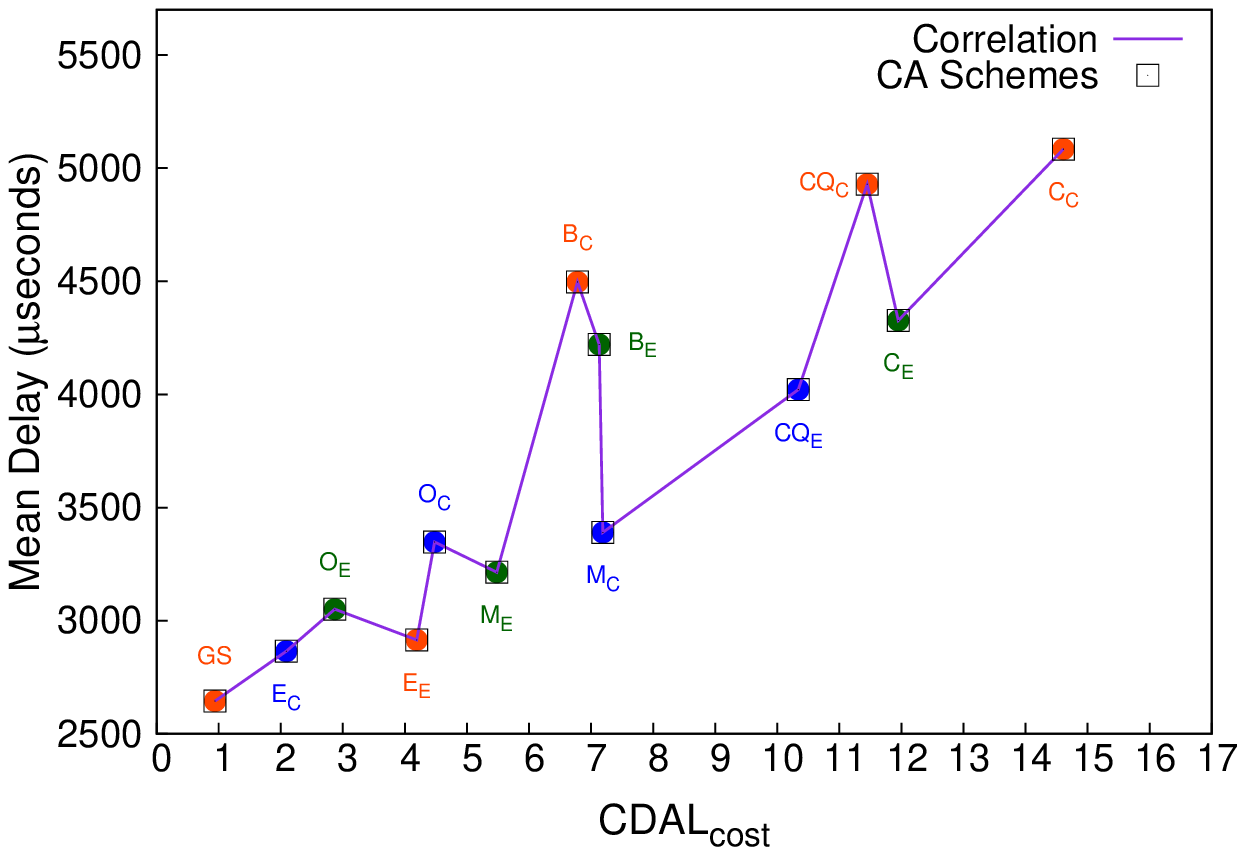}}\hfill%
   \subfloat[CXLS$_{wt}$ vs MD] {\includegraphics[width=.33\linewidth]{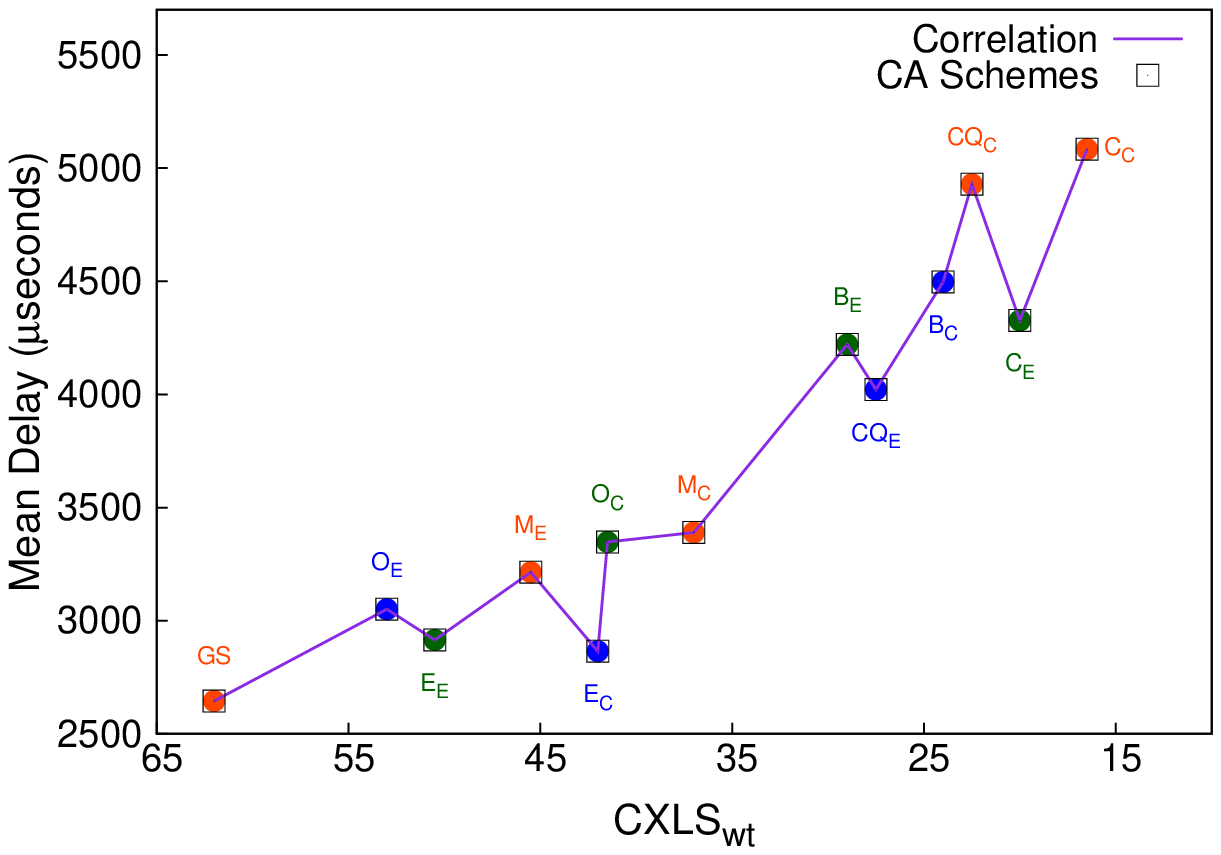}}%
    \end{tabular}
    \caption{Observed correlation of theoretical estimates \& Mean Delay} 
     \label{cMD}
\end{figure*}
\subsection{Results and Analysis}
\begin{figure}[htb!]
                \centering
                \includegraphics[width=8.5cm, height=5cm]{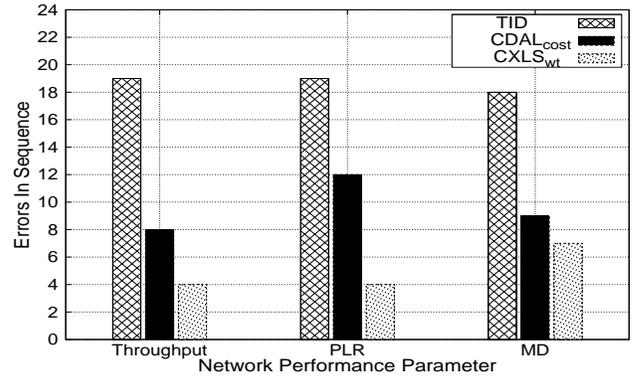}
                \caption{EIS values of CA performance prediction metrics}
                \label{EIS}
        \end{figure}

An exhaustive set of simulations were run for the test-cases described above, and the values of the three performance metrics were recorded. For each CA, we compute the average of the recorded network metric values for all the test-cases to derive, the \textit{average aggregate network throughput} (in Mbps), \textit{average packet loss ratio} (as a $\%$), and  \textit{average mean delay} (in $\mu$seconds). For ease of reference, we henceforth denote them as \textit{Throughput}, \textit{PLR} and \textit{MD}, respectively. For a better representation in result illustrations the CAs are labeled as : BFSCA (B), MISCA (M), CCA (C), CLICA (CQ), OISCA (O), EIZMCA (E) and GSCA (GS). The results are presented in Figures~\ref{cTh},~\ref{cPLR},~\&~\ref{cMD}, through which we demonstrate the observed correlation between theoretical estimates and actual performance metrics. \\
It is necessary to understand the expected correlation of network performance metrics with the prevalent interference. As depicted in Figure~\ref{cor}, the aggregate capacity of a wireless network will deteriorate with rise in the intensity of interference. Further, with increase in the adverse impact of interference, loss of data packets and end to end latency in packet transmission will increase as well. Thus, a reliable theoretical interference estimate must exhibit a similar pattern when plotted against the observed network performance metrics. From Figures~\ref{cTh},~\ref{cPLR},~\&~\ref{cMD}, it can be discerned that TID does not conform to expected correlation and has a haphazard gradient when plotted against network metrics. CDAl$_{cost}$ displays a higher adherence to the expected pattern than TID. CXLS$_{wt}$ estimates exhibit a great similarity to the expected correlation plot gradients. Since all three metrics do not account for the temporal characteristics of wireless communication, a deviation 
from observed patterns is inevitable. Thus, CXLS$_{wt}$ offers the most reliable interference estimates among the three metrics which is visible from the gradients of its plots against the three network performance metrics. We now process and analyze the results to derive the accuracy of each of the three estimation metrics.\\
For every recorded performance metric, we first order the CAs in a \textit{sequence} of increasing performance. In a similar fashion we order the CAs in the increasing order of expected performance, as predicted by the three interference estimation metrics. For both TID and CDAL$_{cost}$, a high estimate implies high interference in the WMN and thereby, a dismal CA performance. In contrast, higher the value of CXLS$_{wt}$, better is the expected performance of the CA \emph{i.e.}, $(Expected \ CA \ Performance \propto CXLS_{wt})$. Thus, the CA sequence in the increasing order of expected performance will be arrived at by orienting CAs in decreasing order of  estimation metric values for both, TID and CDAL$_{cost}$, and in increasing order of estimation metric values for CXLS$_{wt}$.
Next, we compare CA sequences based on experimental data with CA sequences derived from theoretical estimates, to determine the \textit{error in sequence} (EIS) of each prediction metric. Let us consider $n$ CAs which are ordered in a sequence based on the values of a prediction metric. A total of  $\textsuperscript n C_2$ \textit{comparisons} exist between individual CAs in the sequence. These pairwise comparisons of expected CA performances have to be verified against experimental data, by considering the sequence of CAs based on the recorded network metric values as the reference. We determine the total number of comparisons that are \textit{in error} in the CA sequences of theoretical metrics. A comparison in error implies that the expected performance relationship between two CAs as predicted by the estimation metric, is contrary to that observed in actual implementation results. EIS for a particular CA performance prediction metric is the sum of all erroneous comparisons in its CA sequence. Thus, EIS 
is 
a measure of fallacy in the predictions of an estimation metric. Next, we determine the \textit{degree of confidence} (DoC ) which 
represents the level of accuracy that an interference estimation scheme exhibits in its prediction of the performance of a CA. The DoC value for a theoretical estimate is computed through the expression $DoC = (1-(EIS/\textsuperscript n C_2))\times100$, where $n$ is the number of CAs in the sequence.
We elucidate the above procedure through an example. Let us determine the CA sequence in terms of increasing Throughput, which is : \textit{(CCA$_C$ $<$ CLICA$_C$ $<$ CCA$_E$ $<$ CLICA$_E$ $<$ BFSCA$_C$ $<$ BFSCA$_E$ $<$ MISCA$_C$ $<$ OISCA$_C$ $<$ MISCA$_E$ $<$ EIZMCA$_C$ $<$ OISCA$_E$ $<$ EIZMCA$_E$ $<$ GSCA)}. This is the reference ordering of CAs in which CCA is the least efficient and GSCA the best performer in the CA sample set, in terms of observed Throughput. Against this benchmark we compare the CA sequence spawned by CXLS$_{wt}$, which is : \textit{(CCA$_C$ $<$ CCA$_E$ $<$ CLICA$_C$ $<$ BFSCA$_C$ $<$ CLICA$_E$ $<$ BFSCA$_E$ $<$ MISCA$_C$ $<$ OISCA$_C$ $<$ EIZMCA$_C$ $<$ MISCA$_E$ $<$ EIZMCA$_E$ $<$ OISCA$_E$ $<$ GSCA)}. We compare the actual pairwise CA relationships with those predicted by CXLS$_{wt}$ to compute an EIS of $4$  with respect to Throughput. Likewise, the EIS for TID and CDAL$_{cost}$ in terms of Throughput are $19$ and $8$, respectively. EIS for 
all the theoretical estimates with respect to the three observed network metrics are depicted in the Figure~\ref{EIS}. 

Finally, we compute the DoC which is the number of affirmative predictions as a percentage of total number of pairwise comparisons that are possible in the CA sequence ($\textsuperscript {13} C_2$). DoC for the three estimation schemes are presented in Table~\ref{DOC}. 

\begin{table} [h!]
\caption{Performance Evaluation Of Estimation Metrics}
\centering
\tabcolsep=0.10cm
\begin{tabular}{|M{2.5cm}|M{1.7cm}|M{1.7cm}|M{1.7cm}|}
\hline 
\multicolumn{1}{|c|}{\textbf{Performance}}&\multicolumn{3}{|c|}{\textbf{Degree of Confidence} ($\%$)}\\ \cline{2-4}
\multicolumn{1}{|c|}{\textbf{Metric}}&\textbf{TID}&\textbf{CDAL$_{cost}$}&\textbf{CXLS$_{wt}$}\\
\hline  
Throughput&75.64&89.74&94.87\\
\hline 
PLR&75.64&84.61&94.87\\
\hline 
MD&76.92&88.46&91.02\\
\hline  
\end{tabular} 
\label{DOC}
\end{table}

It can be inferred that CXLS$_{wt}$ registers lower EIS than both, TID and CDAL$_{cost}$, in terms of Throughput, PLR and MD. The EIS is halved in CXLS$_{wt}$ estimation in comparison to CDAL$_{cost}$, while it is reduced to almost one fourth when compared to TID estimates.  A similar trend can be observed in the DoC values as well. TID estimates fare worse than both, CDAL$_{cost}$ and CXLS$_{wt}$, as a prediction metric with accuracy levels always below 80\%. CDAL$_{cost}$ exhibits an average performance with DoC values between 80\% and 90\%.  CXLS$_{wt}$ is unarguably the most dependable CA performance prediction metric of the three, as its measure of reliability is always greater than 90\%. 

Further, let us qualitatively assess the prediction patterns of the three estimates. It is discernible that CXLS$_{wt}$ explicitly distinguishes between CAs that will perform well in a WMN and those that will not \emph{eg.}, CXLS$_{wt}$ estimates project  CAs OISCA$_E$, EIZMCA$_E$ \& GSCA as high-performance CAs, the CAs BFSCA$_E$ MISCA$_C$ \& OISCA$_C$ as average performance CAs, and the CAs CCA$_C$ \& CLICA$_C$ as low-performance CAs. These performance predictions are validated by the experimental results. In contrast, TID estimates place the CAs BFSCA$_E$ \& CLICA$_C$ at the bottom of the performance spectrum, and CCA$_C$ as an average-performing CA. Both of the predictions are not in adherence to the actual experimental data. CDAL$_{cost}$ estimates are more accurate than TID, but they fail to compete with CXLS$_{wt}$ as they overlook the spatial aspects of interference alleviation and do not consider the proximity of links that might interfere. Further, all 
the three estimates rightly predict GSCA to be the most efficient CA in the sample set, however only CXLS$_{wt}$ and CDAL$_{cost}$ predict CCA$_C$ to exhibit the poorest performance. 

\section{Conclusions}
Since the problem of interference estimation is NP-hard, the role of a theoretical prediction estimate is limited to exhibit a maximal conformance to the actual recorded behavior of a CA when implemented in a WMN. In this context, CXLS$_{wt}$ proves to be a reliable CA prediction metric with an adherence of over 91\% to actual results, in a fairly extensive sample set of 13 CAs. It does incur a slightly higher computational cost than both, TID estimate and CDAL$_{cost}$, but the overhead of increased algorithmic complexity is adequately compensated by the increase in accuracy levels. Thus, CXLS$_{wt}$ outperforms both TID estimate and CDAL$_{cost}$ as a reliable CA performance prediction metric, which it owes to its spatio-statistical design that ensures a reduced EIS and thus, an enhanced DoC.

\section{Future Work}
Both CDAL$_{cost}$ and CXLS$_{wt}$ are metrics that offer predictions for the whole CA and do not offer an estimate of individual link quality. For a quantitative assessment, such as theoretical upper-bounds of network performance metrics \emph{eg.}, Throughput, a link quality estimate is necessary. Thus we intend to take up this problem and devise a prediction estimate based on the individual link quality. 
\bibliography{ref}

\end{document}